\documentclass[sigconf]{acmart}

\AtBeginDocument{%
  }

\copyrightyear{2022}
\acmYear{2022}
\setcopyright{acmcopyright}
\acmConference[CIKM '22] {Proceedings of the 31st ACM International Conference on Information and Knowledge Management}{October 17--21, 2022}{Atlanta, GA, USA.}
\acmBooktitle{Proceedings of the 31st ACM International Conference on Information and Knowledge Management (CIKM '22), October 17--21, 2022, Atlanta, GA, USA}
\acmPrice{15.00}
\acmISBN{978-1-4503-9236-5/22/10}
\acmDOI{10.1145/3511808.3557404}

\usepackage{times}
\usepackage{soul}
\usepackage{url}
\usepackage{graphicx}
\usepackage{amsmath}
\usepackage{amsthm}
\usepackage{booktabs}
\usepackage{algorithmic}
\usepackage[linesnumbered,ruled,vlined]{algorithm2e}

\usepackage{xspace}
\usepackage{xspace}
\usepackage{subfig}
\usepackage{enumerate}
\usepackage{enumitem}
\usepackage{color}
\usepackage{wasysym}
\usepackage{balance}
\usepackage{multirow}
\usepackage{threeparttable}

\let\oldhat\hat
\renewcommand{\vec}[1]{\mathbf{#1}}
\renewcommand{\hat}[1]{\oldhat{\mathbf{#1}}}
\renewcommand{\matrix}[1]{\mathbf{#1}}

\newcommand{\eg}{\emph{e.g.,}\xspace}

\newcommand{\ie}{\emph{i.e.,}\xspace}

\newcommand{\eat}[1]{}
\newcommand{\paratitle}[1]{\vspace{1ex}\noindent \textbf{#1}}

\begin{document}

\title{Multi-level Contrastive Learning Framework for Sequential Recommendation}

\author{Ziyang Wang}
\authornotemark[1]
\affiliation{%
\institution{CCIIP Laboratory, Huazhong University of Science and Technology}
\institution{Alibaba Group}
\country{China}
}
\email{ziyang1997@hust.edu.cn}

\author{Huoyu Liu}
\authornote{Both authors contributed equally to this research.}
\affiliation{%
  \institution{Alibaba Group}
  \country{China}}
\email{huoyu.lhy@alibaba-inc.com}

\author{Wei Wei}
\authornote{Corresponding Author.}
\affiliation{%
\institution{CCIIP Laboratory, Huazhong University of Science and Technology}
\institution{Joint Laboratory of HUST and Pingan Property \& Casualty Research (HPL)}
  \country{China}
}
\email{weiw@hust.edu.cn}

\author{Yue Hu}
\affiliation{%
  \institution{Alibaba Group}
  \country{China}}
\email{lingshu.hy@alibaba-inc.com}
  
\author{Xian-Ling Mao}
\affiliation{%
  \institution{Beijing Institute of Technology}
  \country{China}
}  
\email{maoxl@bit.edu.cn}
 
\author{Shaojian He}
\affiliation{%
  \institution{Alibaba Group}
  \country{China}}
\email{shaojian.he@alibaba-inc.com}

\author{Rui Fang}
\affiliation{%
  \institution{Ping An Property \& Casualty Insurance company of China, Ltd}
  \country{China}}
\email{fangrui051@pingan.com.cn}
  
 \author{Dangyang Chen}
\affiliation{%
  \institution{Ping An Property \& Casualty Insurance company of China, Ltd}
  \country{China}}
\email{chendangyang273@pingan.com.cn}

\renewcommand{\shortauthors}{Trovato et al.}

\begin{abstract}
Sequential recommendation~(SR) aims to predict the subsequent behaviors of users by understanding their successive historical behaviors.
Recently, some methods for SR are devoted to alleviating the data sparsity problem~(\ie limited supervised signals for training), which take account of contrastive learning to incorporate self-supervised signals into SR. 
Despite their achievements, it is far from enough to learn informative user/item embeddings due to the inadequacy modeling of complex collaborative information and co-action information, such as user-item relation, user-user relation, and item-item relation. 
In this paper, we study the problem of SR and propose a novel multi-level contrastive learning framework for sequential recommendation, named MCLSR.
Different from the previous contrastive learning-based methods for SR, MCLSR learns the representations of users and items through a cross-view contrastive learning paradigm from four specific views at two different levels~(\ie interest- and feature-level). Specifically, the interest-level contrastive mechanism jointly learns the collaborative information with the sequential transition patterns, and the feature-level contrastive mechanism re-observes the relation between users and items via capturing the co-action information (\ie co-occurrence).
Extensive experiments on four real-world datasets show that the proposed MCLSR outperforms the state-of-the-art methods consistently. 
\end{abstract}



\maketitle

\section{Introduction}

Recommendation systems play critical roles in many online services such as E-commerce, video streaming, and music platform due to their success in alleviating the information overload problem. Among these applications, sequential recommendation~(SR) pays attention to the chronological order of users' behaviors and has become a paradigmatic task in recent years.
Given a user behavior history, SR captures the sequential transition patterns among successive items and predicts the next item that the user might be interested in, consistent with many real-world recommendation situations.

The study of the sequential recommendation system is of significant importance and thus has received considerable research interest in recent years.
For instance, there exist several works to treat SR as a sequence modeling task, such as GRU4Rec~\cite{hidasi2015session}, which adopts recurrent neural networks (RNNs) to model the sequential behaviors of users.
Then, SASRec~\cite{kang2018self} uses the self-attention mechanism to capture high-order dynamics from user behavior sequences. 
Further, graph-based methods~\cite{qiu2019rethinking,wang2020global,chang2021sequential} convert each sequence into a graph and model the complex item transitions via graph neural networks~(GNNs).
However, most of these methods are under a supervised learning paradigm and may suffer from the data sparsity problem since their supervision signal is entirely from the observed user behaviors (shown in Fig 1a), which are highly sparse compared to the entire interaction space~\cite{wu2021self}.

\begin{figure}[t]
  \centering
  \includegraphics[width=\columnwidth, angle=0]{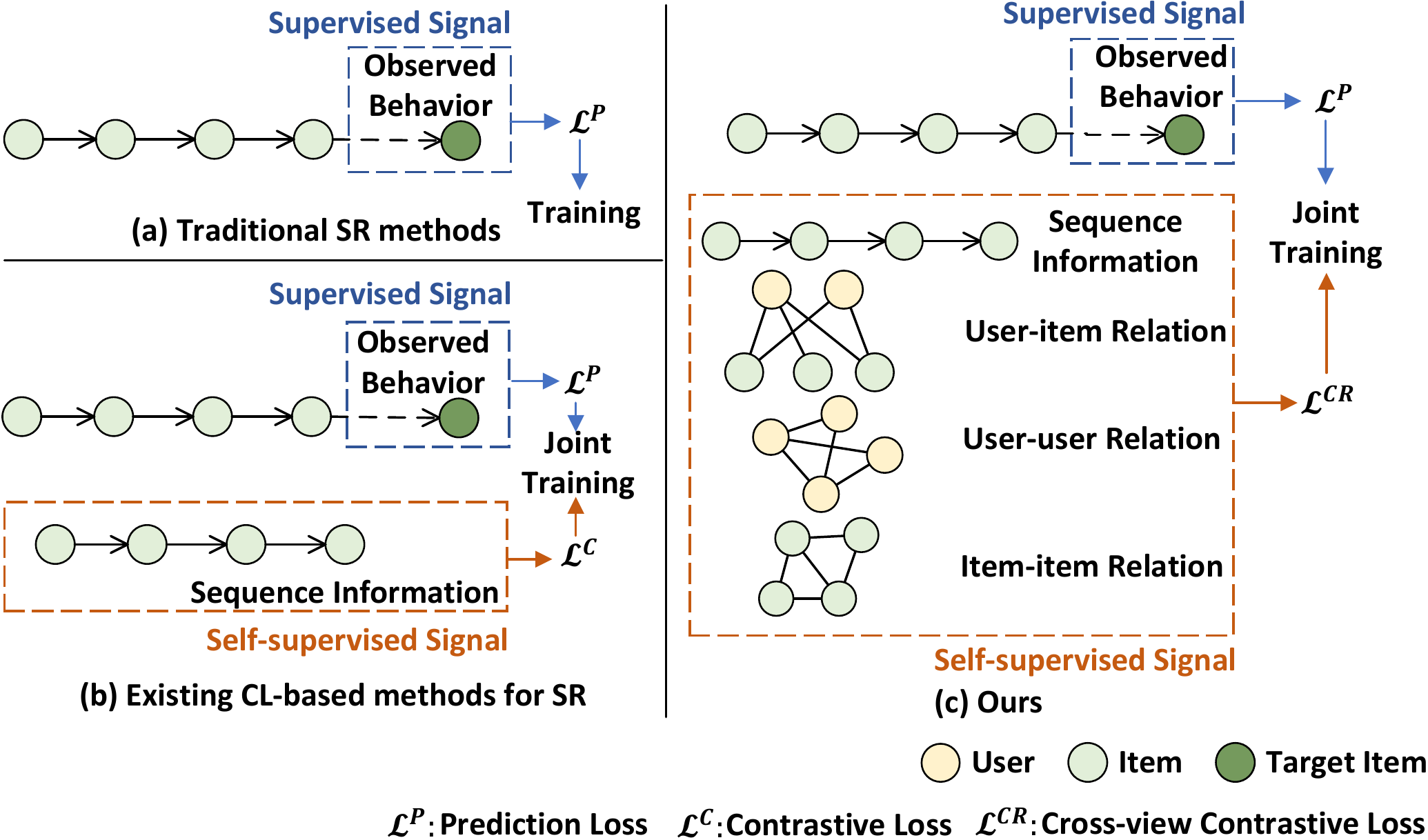}
  \caption{Illustration of training mechanisms of different methods for SR. (a) Traditional methods for SR where the supervised signals are entirely based on the observed user behaviors. (b) Recently contrastive learning-based methods for SR learn the self-supervised signals from the sequence itself. (c) Our proposed method learns rich self-supervised signals by performing cross-view contrastive learning on sequence information, user-item, user-user, and item-item relations.
}
  \label{fig:frameWork}
\end{figure}

Recently, self-supervised learning~(SSL)~\cite{liu2021self} is proposed to mine the supervised signals from the data itself, which shows promising potential to alleviate the data sparsity problem. 
As a typical self-supervised learning technique, contrastive learning~(CL) has gained increasing attention. By extracting the positive and negative samples from the data, contrastive learning aims to maximize the agreement of positive pairs while minimizing the agreement between negative samples. In this way, it can learn discriminative embeddings without explicit extra labels~\cite{wang2021contrastive}.
Based on the principle of contrastive learning, existing CL-based SR methods apply data-level augmentation (S$^3$-Rec~\cite{zhou2020s3} and CL4SRec~\cite{xie2020contrastive}) or model-level augmentation (DuoRec~\cite{qiu2022contrastive}) on user behavior sequence to generate postive and negative pairs, and learn the extra self-supervised signals by contrasting the corresponding pairs.


Despite such achievement, 
the above methods obtain the self-supervised signals entirely from the sequence itself~(shown in Fig 1b), which is insufficient for SR for two reasons.
\emph{First}, since each behavior sequence contains a limited number of items, the self-supervised information obtained from the sequence is inadequate.
\emph{Second}, S$^3$-Rec and CL4SRec generate contrastive pairs by performing simply data augmentation~(\eg item cropping and masking) on behavior sequences, resulting in less information diversity of contrastive pairs, thus the obtained self-supervised signals would be too weak to learn informative embedding~\cite{wang2021contrastive}. Due to the insufficiency of directly exploiting contrastive learning on sequential views (i.e. user behavior sequences), it motivates us to explore more views and generate more informative pairs for contrastive learning. However, it is non-trivial to define an appropriate contrastive learning framework with more contrasting views for SR, which requires us to address the following fundamental issues:
(1) \emph{How to select proper views for contrastive learning}: As mentioned above, more views are desired for the contrastive learning of SR. An essential requirement is that the selected views should be informative and can reflect user preferences.
In fact, collaborative information~\cite{wang2019neural} (\eg user-item relation) and co-action information~\cite{gao2018bine} (\eg user-user relation and item-item relation) are two significant factors for user preference learning, which show strong potential to help obtain rich self-supervised signals and should be carefully considered.
(2) \emph{How to set a proper contrastive task}: Proper design of contrastive tasks is critical for contrastive learning~\cite{wang2021contrastive}. In general, similar contrastive views would make the self-supervised signals too weak to learn informative embeddings.
Therefore, it is important to ensure a clear diversity of information between the contrasting views.

In light of the aforementioned limitations and challenges, in this paper, we propose a \textbf{m}ulti-level \textbf{c}ontrastive \textbf{l}earning framework for \textbf{s}equential \textbf{r}ecommendation (MCLSR). To effectively learn the self-supervised signals, except for the sequential view, we construct three graph views and adopt a multi-level cross-view contrastive mechanism to learn collaborative information, co-action information and sequential transitions (shown in Fig 1c). Specifically, four views of SR are given firstly, \ie sequential view, user-item view, user-user view, and item-item view. 
Then MCLSR performs a cross-view contrastive learning paradigm on two levels (\ie, interest- and feature-level).
At the interest-level, MCLSR obtains the sequential transition patterns from the sequential view and the collaborative information from the user-item view, where the contrastive mechanism is performed to capture the complementary information between the two views.
At the feature-level, MCLSR re-observed the relation between users and items via performing GNNs on the user-user view and item-item view. By applying contrastive learning to learn discriminative information on two views, MCLSR can capture the self-supervised signals from the co-action information between users~(items) to further enhance representation learning.

To summarize, this work makes the following main contributions:
\begin{itemize}[leftmargin=*]
\item We exploit contrastive learning with collaborative information and co-action information to alleviate the data sparsity problem in studying the sequential recommendation task. Towards this end, we propose a novel recommendation framework that captures sequential transition patterns, collaborative signals and co-action signals from four specific views.
\item The proposed MCLSR performs cross-view contrastive learning at interest- and feature-level. The former learns self-supervised signals from collaborative information and sequential transition patterns, and the latter captures the co-action information to learn informative user/item embeddings.
\item We conduct extensive experiments on four real-world datasets, and the results demonstrate the superiority of MCLSR and the effectiveness of each key component.
\end{itemize}

\section{Related Work}

In this section, we will briefly review several lines of work closely
related to ours, including sequential recommendation and contrastive learning.

\subsection{Sequential Recommendation} 
Compared with session-based recommendation~\cite{wu2019session}, sequential recommendation usually considers user ID and their behavior sequence in a longer time period.
Most of the early attempts for SRS are based on Markov Chain, which infers a user’s next action based on the previous one. 
For example, FPMC~\cite{rendle2010factorizing} captures the sequential patterns by first-order Markov Chain, which is then extend to higher order Markov Chain~\cite{he2016fusing}.
%
To capture long-term and multi-level cascading dependencies, deep learning techniques are introduced into SRS. 
For instance, RNN-based methods~\cite{hidasi2015session, wu2017recurrent,xu2019recurrent} regard SRS as a sequential modeling problem and apply recurrent neural networks to capture the sequential transition patterns. 
Further, CNN-based methods~\cite{tang2018personalized,yuan2019simple} treat each sequence as an image and adopt convolution networks to model the union-level sequential patterns.
Then some advanced techniques are incorporated into SRS, such as self-attention attention network~\cite{kang2018self,wang2018attention,sun2019bert4rec,ma2019hierarchical,li2020time}, memory network~\cite{chen2018sequential,huang2018improving,zhang2021learning,cai2021category}, capsule network~\cite{li2019multi,cen2020controllable} and graph neural networks~\cite{xu2019graph,ma2020memory,chang2021sequential,zhang2022dynamic}.
Typically, SASRec~\cite{kang2018self} stacks multi-head self-attention blocks to learn dynamic item transition patterns. 
MIND~\cite{li2019multi} leverages dynamic routing to obtain multiple interests of users.
MA-GNN~\cite{ma2020memory} proposes a memory augmented graph neural network to capture both items’ short-term contextual information and long-range dependencies for sequential recommendation.
However, the above methods mainly focus on the modeling of sequential transition in a supervised paradigm, where the supervised signals are entirely based on the observed user behaviors. Due to the limited observed user behaviors, the above methods face the problem of data sparsity. In this paper, we mainly focus on employing a multi-level cross-view contrastive learning paradigm to alleviate the data sparsity problem.

\subsection{Contrastive Learning}
The main idea of contrastive learning is to learn informative representations by contrasting positive pairs against negative pairs, which shows impressive achievement in visual representation learning~\cite{chen2020simple}, natural language process~\cite{yang2019reducing,zhu2021contrastive}, and graph neural networks~\cite{you2020graph,jiang2021contrastive,chu2021cuco}.

Recently, some studies are proposed to introduce contrastive learning into recommendation system~\cite{ma2020disentangled,xia2021graph,xia2021self,qin2021world,qiu2021memory,yu2021socially,yu2021graph,zhao22AAAI,zou2022multi}. For instance, SGL~\cite{wu2021self} provides an auxiliary signal for existing GCN-based recommendation models by taking node self-discrimination as the self-supervised task.
SEPT~\cite{yu2021socially} designs a socially aware self-supervised framework for learning discrimination signals from the user-item graph and social graph. 
Some efforts also introduce contrastive learning into sequential recommendation~\cite{zhou2020s3,qiu2022contrastive,chen2022intent}.
S$^{3}$-Rec~\cite{zhou2020s3} devises four auxiliary self-supervised objectives for data representation learning by using the mutual information maximization.
CL4SRec~\cite{xie2020contrastive} applies three data augmentation~(\ie crop, mask and reorder) to generate positive pairs, and contrasts positive pairs to learn robust sequential transition patterns.
DuoRec~\cite{qiu2022contrastive} proposes a dropout-based model-level augmentation model with a supervised positive sampling strategy to capture the self-supervised signal from the sequence.
Despite the achievement, the above contrastive learning-based methods for SR mainly focus on learning the self-supervised signals from each sequence.
However, due to the limited information within the sequence, the obtained self-supervised signal will be too weak to learn informative embedding.

\section{Preliminary}
In this section, we first formulate the problem of sequential recommendation, then we introduce the construction process of three graph views and the architecture of the graph encoder layer.

\subsection{Problem Formulation}
Assume we have a set of users $u \in \boldsymbol{\mathcal{U}}$ and a set of of items $v \in \boldsymbol{\mathcal{V}}$. For each user, $\vec{S}^{(u)} = \{ v^{(u)}_1, v^{(u)}_2, \cdots, v^{(u)}_{|\mathcal{S}|} \}$ denotes the sequence of user historical behaviors in chronological order, where $v^{(u)}_j$ denotes the $j^{th}$ item interacted by the user.
Given an observed sequence $\vec{S}^{(u)}$, the typical task of sequential recommendation is to predict the next items that the user $u$ is most likely to be interacted with.

\subsection{Graph Construction}

An item can be involved in multiple user behavior sequences, from where we can obtain useful collaborative information~\cite{wang2019neural} and co-action information~\cite{gao2018bine}. Thus extra graph views are constructed here to explore the collaborative signals and co-action signals for SRS.
Based on the users' historical behavior sequences, we first obtain an user-item interaction matrix $\boldsymbol{\mathcal{M}^{uv}} \in \mathbb{R}^{|\boldsymbol{\mathcal{U}}| \times |\boldsymbol{\mathcal{V}}|}$, where $\boldsymbol{\mathcal{M}^{uv}_{ij}} > 0$ denotes that item $j$ is appeared in user $i$'s behavior sequence $\vec{S}^{i}$~(\ie $v_j \in \vec{S}^{i} $) and 0, otherwise.

\paratitle{User-item graph.}
The user-item graph is a typical bipartite graph, which is constructed by aggregating cross-user behavior sequences.
Let 
\begin{math}
  \mathcal{G}^{uv} = ( \mathcal{V}^{uv}, \mathcal{E}^{uv} )
\end{math}
be the user-item graph, where $\mathcal{V}^{uv}$ denotes the node set of graph $\mathcal{G}^{uv}$ that contains all users in $\boldsymbol{\mathcal{U}}$ and all items in $\boldsymbol{\mathcal{V}}$, and $\mathcal{E}^{uv} = \{ e^{uv}_{ij} = \boldsymbol{\mathcal{M}^{uv}_{ij}} | \boldsymbol{\mathcal{M}^{uv}_{ij}} > 0 \}$ indicates the edge set of graph $\mathcal{G}^{uv}$ that contains user-item interactions, where the weight of edge $e^{uv}_{ij}$  represents the number of user $i$ interacts with item $j$.

\paratitle{User-user/item-item graph.}
The user-user~(item-item) graph is constructed to explore the co-action signals between users~(items). 
Based on the interaction matrix $\boldsymbol{\mathcal{M}^{uv}}$, we can obtain a user-user matrix\footnote{Here we present how to construct the user-user graph $\mathcal{G}^{uu}$, and the item-item graph $\mathcal{G}^{vv}$ can be constructed similarly.}
$
\boldsymbol{\mathcal{M}^{uu}} = 
\left( \boldsymbol{\mathcal{M}^{uv}} \right)
\left( \boldsymbol{\mathcal{M}^{uv}} \right)^T
$.
Let
\begin{math}
  \mathcal{G}^{uu} = ( \mathcal{V}^{uu}, \mathcal{E}^{uu} )
\end{math}
be the user-user graph, where $\mathcal{V}^{uu}$ denotes the graph node set that contains all users in $\boldsymbol{\mathcal{U}}$, and $\mathcal{E}^{uu} = \{ e^{uu}_{ij} = \boldsymbol{\mathcal{M}^{uu}_{ij}} | \boldsymbol{\mathcal{M}^{uu}_{ij}} > 0 \}$ indicates graph edge set that contains co-action information, where the weight of each edge denotes the number of co-action behaviors between user $i$ and user $j$.

\subsection{Graph Encoder Layer}
To fully exploit the collaborative information and co-action information from the graphs, a specific graph encoder layer is employed here to extract the node features. Due to the effectiveness and lightweight architecture of LightGCN~\cite{he2020lightgcn}, we employ its message propagation strategy to encode the node features: 
\begin{equation}
\begin{split}
    \matrix{X}^{(l)} = \text{GraphEncoder}(\matrix{X}, \matrix{A})
    = \matrix{D^{-\frac{1}{2}} A D^{-\frac{1}{2}}} \matrix{X}^{(l-1)}, \\
\end{split}
\label{eq:gnn_layer}
\end{equation}
where $\matrix{A}$ indicates the adjacency matrix of the graph and $\matrix{D}_{ii} = \sum_{j=0} \matrix{A}_{ij}$ denotes the corresponding diagonal degree matrix. $(l)$ indicates the depth of graph encoder layers, $\matrix{X}^{(0)}$ indicates the input node features and $\matrix{X}^{(l)}$ is the output of the graph encoder layer.


\section{Method}

\begin{figure*}[t]
  \centering
  \includegraphics[width=1.9\columnwidth, angle=0]{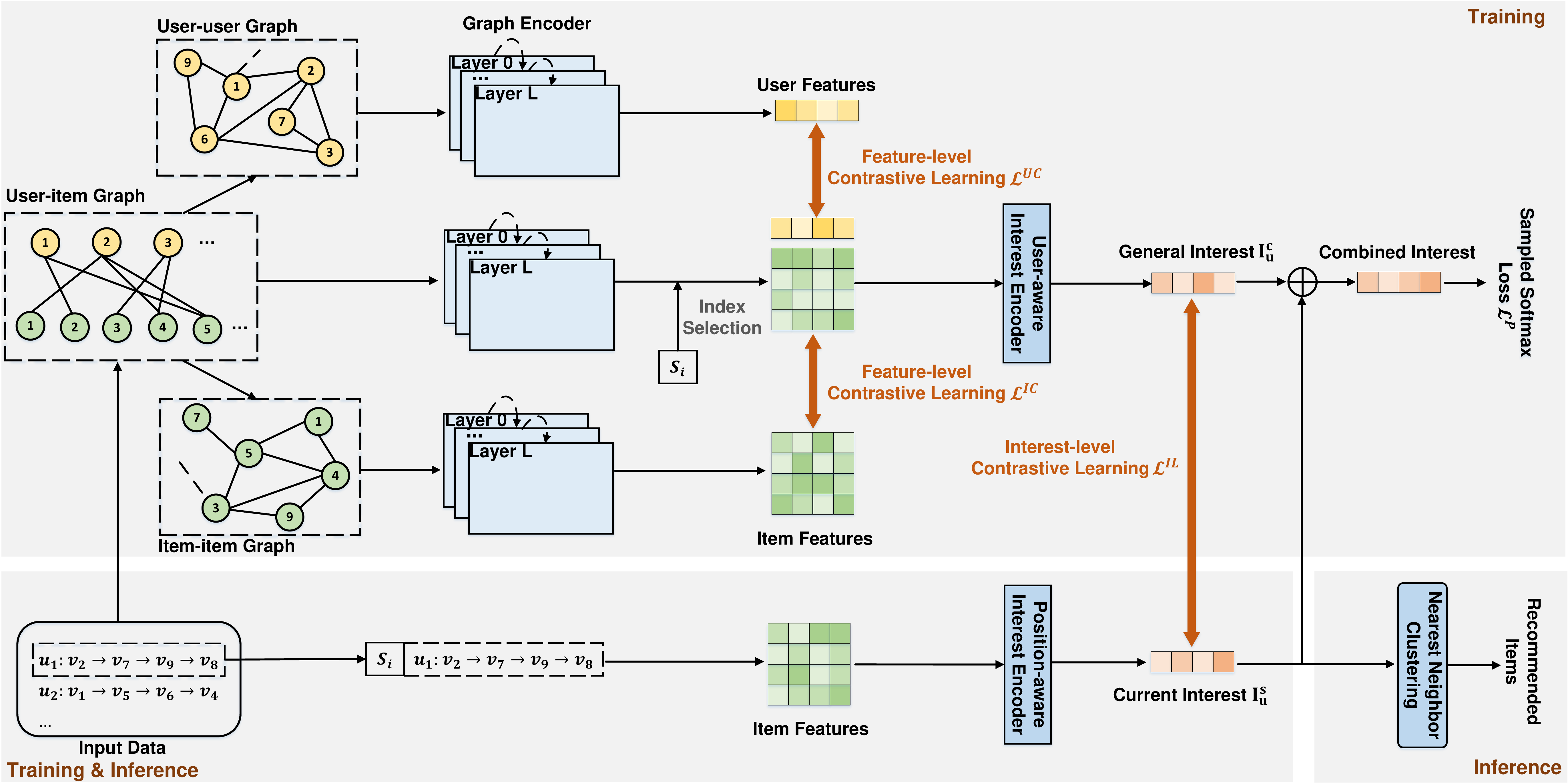}
  \caption{An overview of the proposed framework. $\oplus$ denotes the element-wise summation. 
}
  \label{fig:frameWork}
\end{figure*}
The overview of the proposed multi-level contrastive framework is presented in Figure \ref{fig:frameWork}. which comprises four main components:
1) \emph{Graph construction layer.} It constructs user-item, user-user and item-item graphs via aggregating the user behavior sequences;
2) \emph{Interest-level contrastive learning layer.} It first learns the current interest of the user from user behavior sequences and the general interest of the user from the user-item graph, then a cross-view contrastive mechanism is performed.
3) \emph{Feature-level contrastive learning layer.} It obtains the user and item features from the user-item, user-user and item-item graphs, and then it performs cross-view contrastive learning;
4) \emph{Joint training.} It jointly optimizes the prediction loss, interest- and feature-level contrastive loss to update the model parameters. 
In the following sections, we will present the technical details of MCLSR.

Here, we first construct a user embedding matrix $\matrix{H}^{u} \in \mathcal{R}^{|\boldsymbol{\mathcal{U}}| \times d}$ and an item embedding matrix $\matrix{H}^{v} \in \mathcal{R}^{|\boldsymbol{\mathcal{V}}| \times d}$, where $d$ is the dimension of the embedding. 
The input of our model is the user behavior sequence $\vec{S}^{(u)} = \{ v^{(u)}_1, v^{(u)}_2, \cdots, v^{(u)}_{|\mathcal{S}|} \}$, which is fed into a specific embedding layer and transformed to the user embedding $\vec{h}^u \in \mathbb{R}^{d}$ and the corresponding item embedding matrix $\matrix{E}^{u} = [ \vec{h}^v_{1}, \vec{h}^v_{2}, \cdots, \vec{h}^v_{n} ]$.

\subsection{Interest-level Contrastive Learning}

Different from previous SR studies~\cite{chen2018sequential} that mainly focus on the transition patterns on the current sequence, we aim to introduce collaborative information into SR to capture the general preferences of users. Then by applying contrastive mechanism, the extra self-supervised signals from the complementary information between sequential transition patterns and collaborative information is learned to alleviate the data sparsity problem.

\paratitle{Current interest learning.}
This subsection aims to capture the users' preferences from the user behavior sequences (\ie sequential view). Since different items have distinct importance for current prediction, a self-attention mechanism~\cite{lin2017structured} is applied to model the user behavior sequences. 
Given the item embedding matrix $\matrix{E}^u$, we first use a trainable position matrix to incorporate the sequential order information into sequence, \ie $\matrix{E}^{u,p} = [\vec{h}^v_{1} + \vec{p}_1, \vec{h}^v_{2} + \vec{p}_2, \cdots, \vec{h}^v_{n} + \vec{p}_n]$.
Then an attention matrix $\matrix{A}^{s}$ is computed as follows:
\begin{equation}
    \matrix{A}^{s} = \text{softmax} \left( \matrix{W}_2 \text{tanh} ( \matrix{W}_1 (\matrix{E}^{u,p})^T)\right),
\end{equation}
where $\matrix{W}_1 \in \mathbb{R}^{4d \times d}$ and $\matrix{W}_2 \in \mathbb{R}^{4d}$ are trainable parameters. The dimension of the output matrix $\matrix{A}^{s}$ is $ \mathbb{R}^{n}$, where each element $\matrix{A}_{j}^{s}$ denotes the affinity between the user preference and $j^{th}$ item in the user behavior sequence. Finally, the user preferences from the sequential view (named current interest) can be obtained by:
\begin{equation}
    \matrix{I}^s_u =  \matrix{A}^s \matrix{E}^u.
\end{equation}

\paratitle{General interest learning.} 
To fully explore the collaborative information, here we frame the user interests from the cross-user interaction information in the user-item graph $\mathcal{G}^{uv}$ . 
To obtain the user features and item features, a graph encoder layer is employed as follows\footnote{In the equation, we use $\mathcal{G}^{*}$ to represent the adjacency matrix of the graph for ease of reading.}:
\begin{equation}
\begin{split}
    \matrix{H}^{all,uv} &= \text{GraphEncoder}^{(l)} (\matrix{H}^{all}, \mathcal{G}^{uv}),
\end{split}
\label{eq:GE1}
\end{equation}
where $\matrix{H}^{all} \in \mathbb{R}^{(|\boldsymbol{\mathcal{U}}| + |\boldsymbol{\mathcal{V}}|) \times d}$ is the initial node feature matrix that contains user and item features (\ie $\matrix{H}^{all}  = [\matrix{H}^{u} || \matrix{H}^{v}]$, where $||$ denotes concatenation operation), and \text{GraphEncoder} indicates the graph encoder layer that defined in Equation (\ref{eq:gnn_layer}). $\matrix{H}^{all,uv} \in \mathbb{R}^{ (|\boldsymbol{\mathcal{U}}| + |\boldsymbol{\mathcal{V}}|) \times d}$ is the learned node feature matrix from user-item graph. 

Then for a given user $u$ and corresponding behavior sequence $\vec{S}^u$, we can obtain the corresponding user embedding $\vec{h}^{u, uv} \in \mathbb{R}^{d}$ and item embedding matrix $\matrix{E}^{u, uv} \in \mathbb{R}^{n \times d}$ by index selection from the learned node feature matrix $\matrix{H}^{all,uv} \in \mathbb{R}^{(|\boldsymbol{\mathcal{U}}| + |\boldsymbol{\mathcal{V}}|) \times d}$ according to the indices of user and items.
To estimate the importance of each item based on the user preference, an attention matrix is computed based on the user features and item features::
\begin{equation}
    \matrix{A}^{c} = \text{softmax} \left( \text{tanh} \left( \matrix{W}_3 \vec{h}^{u, uv} \right) \left( {\matrix{E}^{u, uv}}\right)^T \right),
\end{equation}
where $\matrix{W}_3 \in \mathbb{R}^{d \times d}$ is a trainable transform weight, $\matrix{A}^{c} \in \mathbb{R}^{n}$ is the attention matrix between user preference and items. Then the user preferences from the user-item view (named general interest) can be obtained as follows:
\begin{equation}
    \matrix{I}^{c}_u = \matrix{A}^{c} {\matrix{E}^{u, uv}}.
\end{equation}

\paratitle{Cross-view contrastive learning.}
To learn the complementary information from sequential transition patterns and collaborative information, it is meaningful to perform the contrastive learning on the the sequential view (current preference $\matrix{I}_u^{s}$) and user-item view (general preference $\matrix{I}_u^{c}$).
Here, we first feed $\matrix{I}_u^{s}$ and $\matrix{I}_u^{c}$ into a multi-layer perceptron~(MLP) to project them into the space where contrastive loss is calculated:
\begin{equation}
\begin{split}
    \matrix{T}^{I,s} &= \left(\matrix{W}^p_2  \sigma (\matrix{W}^p_1 \matrix{I}_{u}^{s} + \vec{b}^{p}_1) + \vec{b}^{p}_2 \right),  \\
    \matrix{T}^{I,c} &= \left(\matrix{W}^p_2  \sigma (\matrix{W}^p_1 \matrix{I}_{u}^{c} + \vec{b}^{p}_1) + \vec{b}^{p}_2 \right),  
\end{split}
\end{equation}
where $\matrix{W}^p_* \in \mathbb{R}^{d \times d}, \vec{b}^p_* \in \mathbb{R}^{d}$ are trainable parameters and $\sigma$ denotes \text{ELU} non-linear activation function.

To learn the self-supervised signal from two views, it is essential to define the positive and negative samples for the interest-level contrastive mechanism. Inspired by the work of contrastive learning in graph neural networks~\cite{wang2021contrastive}, we take the interests of the same user from two views~(\ie sequential view and user-item view) as a pair of positive samples. Moreover, we naturally treat the interests of different users as pairs of negative samples. Then the interest-level contrastive 
loss can be computed as follows:
\begin{equation}
\mathcal{L}^{IL} = \sum_{i=1} - \text{log} 
                  \frac{ \Psi \left( \matrix{T}^{I,s}_{i}, \matrix{T}^{I,c}_{i} \right)}
                  {
                  \sum \limits_{j} \Psi \left( \matrix{T}^{I,s}_{i}, \matrix{T}^{I,c}_{j} \right) + 
                  \sum \limits_{j \neq i} \Psi \left( \matrix{T}^{I,s}_{i}, \matrix{T}^{I,s}_{j}  \right)
                  },
\end{equation}
where $\Psi$ denotes $\text{exp}\left( \text{sim} ( \cdot, \cdot) / \tau \right)$, $\text{sim} (\cdot, \cdot)$ denotes the cosine similarity function and $\tau$ is a temperature parameter. 

\subsection{Feature-level Contrastive Learning}
Directly exploring the user-item graph is insufficient to capture the co-action information between users~(items). In fact, the co-action information~\cite{gao2018bine} is essential for measuring user-user~(item-item) relationships and learning the user preferences. Thus a user-user~(item-item) graph is constructed to effectively capture the co-action signals between users~(items). 
For each user\footnote{Here we show the \emph{feature-level contrastive learning} for user features, and the process for item features is similarly.}, we learn the user features from both the user-item view and user-user view, where the contrastive mechanism is performed to learn self-supervised signals by capturing the discriminative information from two graph views and complement each other.

\paratitle{Feature learning.} 
To obtain the collaborative information and co-action information, we first extract the user features from user-item view and user-user view, where a graph encoder layer is applied:
\begin{equation}
\begin{split}
    [\matrix{H}^{u, uv}||\matrix{H}^{v,uv}] &= \text{GraphEncoder}^{(l)} ([\matrix{H}^{u}||\matrix{H}^{v}], \mathcal{G}^{uv}), \\
    \matrix{H}^{u, uu} &= \text{GraphEncoder}^{(l)} (\matrix{H}^{u}, \mathcal{G}^{uu}),
\end{split}
\label{eq:GE2}
\end{equation}
where $\matrix{H}^{u, uv}, \matrix{H}^{u, uu}$ denotes the user features obtained from user-item graph $\mathcal{G}^{uv}$ and user-user graph $\mathcal{G}^{uu}$, respectively.
Note that the weight of edge in $\mathcal{G}^{uu}$ denotes the number of co-action, which means high co-action pairs show a more critical influence during graph propagation. 

\paratitle{Cross-view contrastive learning.} 
Then the obtained user features from two graphs are fed into an MLP and projected into the space where contrastive loss is calculated:
\begin{equation}
\begin{split}
    \matrix{T}^{F,uu} &= \matrix{W}^p_4  \sigma (\matrix{W}^p_3 \matrix{H}^{u, uu} + \vec{b}^{p}_3) + \vec{b}^{p}_4,  \\
    \matrix{T}^{F,uv} &= \matrix{W}^p_4  \sigma (\matrix{W}^p_3 \matrix{H}^{u, uv} + \vec{b}^{p}_3) + \vec{b}^{p}_4,
\end{split}
\end{equation}
where $\matrix{W}^p_* \in \mathbb{R}^{d \times d}$, $\vec{b}^p_* \in \mathbb{R}^{d}$ are trainable parameters. 

Considering that each user is involved in two graph views, where we can capture the user-item collaborative information and user-user co-action information, respectively. 
To capture the complementary information between two graph views and obtain discriminative user features, we naturally treat the features of the same user obtained in two graph views as a pair of positive sample while the features of different users as pairs of negative samples:
{
\begin{equation}
    \mathcal{L}^{UC} = \sum_{i=1} - \text{log} 
                      \frac{ \Psi \left( \matrix{T}^{F,uv}_{i},\matrix{T}^{F,uu}_{i} \right)}
                      {
                      \sum \limits_{j} \Psi \left(  \matrix{T}^{F,uv}_{i}, \matrix{T}^{F,uu}_{j} \right) + 
                      \sum \limits_{j \neq i} \Psi \left( \matrix{T}^{F,uv}_{i}, \matrix{T}^{F,uv}_{j} \right)
                      },
\end{equation}
}
where $\mathcal{L}^{UC}$ denotes the contrastive loss for user features and the contrastive loss for item features $\mathcal{L}^{IC}$ can be calculated in a similar way.
The final feature-level contrastive loss $\mathcal{L}^{FL}$ is computed as follows: 
\begin{equation}
\mathcal{L}^{FL} = \mathcal{L}^{UC} + \mathcal{L}^{IC}.    
\end{equation}

\subsection{Training and Inference}
\paratitle{Training phase.}
After computing the user interest representations from sequential view and user-item view, we sum them up to obtain the combined user interest representations:
\begin{equation}
    \matrix{I}^{comb}_u = \alpha \matrix{I}^{s}_u + (1-\alpha) \matrix{I}^{c}_u, 
\label{eq:finalInterest}
\end{equation}
where $\alpha$ is a trade-off hyper-parameter. Given a training sample (u, o) with the user interest embedding $\matrix{I}^{comb}_u$ and target embedding $\vec{h}^{v}_o$, the likelihood of the user $u$ interacting with the item $o$ can be computed by sampled softmax method. Furthermore, the objective function for prediction is to minimize the following negative log-likelihood:
\begin{equation}
    \mathcal{L}^{p} = \sum_{u \in U} - log \frac{\exp((\matrix{I}^{comb}_u)^T  \vec{h}^{v}_o)}
                      {\sum_{k \in Sample (\boldsymbol{\mathcal{V}})}\exp ( (\matrix{I}^{comb}_u)^T  \vec{h}^{v}_k)}.
\end{equation}
The overall objective is given as follows:
\begin{equation}
    \mathcal{J}(\theta) = \mathcal{L}^{p} + \beta \mathcal{L}^{IL} + \gamma \mathcal{L}^{FL},
\end{equation}
while $\beta$ and $\gamma$ are trade-off hyper-parameters. Noted that, we jointly optimize the three throughout the training.

\paratitle{Inference phase.}
For the inference phase, we use the current interest $\matrix{I}^{s}_u$ to perform downstream tasks because:
i) To avoid the problem of information leakage, we only use the training data to construct three graphs during training and inference, so the general interest of users cannot be generated during inference.
ii) After optimizing $\mathcal{J}(\theta)$, the collaborative information and co-action information are learned in the user and item embeddings, thus it is enough to use the current interest $\matrix{I}^{s}_u$ to perform downstream tasks.
Then the candidate items are clustered based on the inner product:
\begin{equation}
    R(u, N) = \text{Top-N}_{v \in V} \left( (\matrix{I}^{s}_u)^T \vec{h}_v \right),
\end{equation}
where $R(u, N)$ denotes the top-N items to be recommended.

\section{Experiment}








{
\renewcommand\arraystretch{1}
\begin{table}[t]
    \setlength{\tabcolsep}{3pt}
	\centering
	\small
	\caption{Statistics of the used datasets.}
	\begin{tabular}{lccccc}
	\toprule[1pt]
		{ \text{Dataset} } & \text{\# user} & \text{\# item} & \text{\# interactions} & \text{Avg. len.} & \text{Sparsity}\\
		\hline
		{ \text{Books} }  & 459,133 & 313,966 & 8,898,041 & 9.7 & 99.993\% \\
 		{ \text{Clothing} } & 39,387 & 23,034 & 278,677 & 6.9 & 99.969\% \\
		{ \text{Toys} } & 75,258 & 64,444 & 1,097,592 & 9.6 & 99.977\% \\
		{ \text{Gowalla} } & 65,506 & 174,606 & 2,061,264 & 14.5 & 99.982\% \\
	\bottomrule[0.8pt]
\end{tabular}
\label{tab:datasets}
\end{table}
}

\subsection{Experimental Settings}

\paratitle{Datasets.} 
We conduct experiments on four public datasets.

\begin{itemize}
    \item \textbf{Amazon}\footnote{\url{http://jmcauley.ucsd.edu/data/amazon/}} consists of product reviews and metadata from \url{Amazon.com}~\cite{mcauley2015image}, and in this study we choose three representative categories: \textbf{Books}, \textbf{Clothing} and \textbf{Toys}. 
    
    \item \textbf{Gowalla}\footnote{\url{https://snap.stanford.edu/data/loc-gowalla.html}} is a widely used check-in dataset which is from a well-known location-based social networking website.
    
\end{itemize}

Following~\cite{cen2020controllable}, we remove the items that appear less than five times, and the max length of each training sample is set to 20. 
The users of each dataset are split into training, validation, and test sets by the proportion of 8:1:1. The model is trained on the entire click sequences of training users. During the training phase, we incorporate a commonly used set of training sequential recommendation models. In detail, we view each item in the user interaction sequence as a potential target item, where the behaviors happen before the target item is used to generate the users' interest representation. During the inference phase, we choose to generate the users' interest representation from the first 80\% of user behaviors and compute the evaluation metric by predicting the remaining 20\% of user behaviors by following~\cite{cen2020controllable}.
The statistics of datasets, after preprocessing, are shown in Table \ref{tab:datasets}.

\paratitle{Baselines.} To fully evaluate the performance of our method for SR, we compare our method with classic methods as well as state-of-the-art methods.
\begin{itemize}[leftmargin=*]
    \item \textbf{Pop} directly recommends top-$N$ popular items in the training data during inference.
    \item \textbf{GRU4Rec}~\cite{hidasi2015session} is the first work that applies recurrent neural network for SR.
    \item \textbf{SASRec}~\cite{kang2018self} stacks several multi-head self-attention blocks to capture the sequential transition patterns.
    \item \textbf{ComiRec-SA}~\cite{cen2020controllable} proposes a multi-interest framework for SR by employing a multi-head self-attention network, where different heads correspond to different interests of users.  
    \item \textbf{GCSAN}~\cite{xu2019graph} combines the graph neural network and self-attention mechanism to learn both short- and long-term dependencies between items.
    \item \textbf{S$^3$-Rec$_{\text{MIP}}$}~\cite{zhou2020s3} utilizing the mutual information maximization (MIM) principle to extract the self-supervised signals from the item transitions.
    \item \textbf{CL4SRec}~\cite{xie2020contrastive} utilizes contrastive mechanism with data augmentation to learn discriminative information.
    \item \textbf{DuoRec}~\cite{qiu2022contrastive} proposes a model-level augmentation method with a  positive sampling strategy to capture the self-supervised signal from the user behavior sequences.
\end{itemize}


\paratitle{Evaluation metrics.} 
Following previous work~\cite{cen2020controllable,xie2020contrastive}, we adopt three widely used ranking-based metrics for sequential recommendation: Recall@N, NDCG@N, and Hit@N. 
Recall@N denotes the proportion of ground truth items included in the top-N recommended list, NDCG@N measures the positions of recommended items and evaluates the ranking quality of the recommended list, and Hit@N denotes the percentage that the top-N recommended list contains at least one ground truth item. 

\paratitle{Implementation details.}
For a fair comparison, all methods are optimized with Adam optimizer with a learning rate of 0.001. The embedding size is set to 64, and the mini-batch size is set to 128. The number of negative samples for sampled softmax loss is set to 1280. For baselines with Transformer blocks (\eg SASRec, CL4SRec, and DuoRec), we select the number of Transformer blocks in $\{1, 2, 3\}$ and select the dropout ratio in $\{0.1, 0.2, ..., 0.9\}$ in the validation set.
For other parameters of baseline methods, we follow the settings given by the original papers if they had provided and otherwise we perform a grid search in the validation set.
For our method, the depth of GNN layer is set to 2 selected from $\{0, 1, 2, 3\}$ in the validation set, and the temperature $\tau$ is set to 0.5. The trade-off parameters $\{\alpha, \beta, \gamma \}$ are set to $\{0.5, 1, 0.05\}$ for all datasets. 
To decrease the noise and reduce the computational complexity, the neighborhood number of each node on the user-user and item-item graph is set to 50 by filtering edges with small weights.

{
\renewcommand{\arraystretch}{1}
\begin{table*}[]
\center
\small
\setlength{\tabcolsep}{2.5pt}
\caption{Effectiveness comparison between MCLSR and state-of-the-art approaches. $^\dagger$  denotes the performance improvement over the best baseline is statistically significant with p-value $< 0.01$.}
\begin{tabular}{clcccccccccc}
\toprule[0.8pt]
Datasets                & Metric & POPRec & GRU4Rec & SASRec & ComiRec-SA & GCSAN & S$^3$-Rec$_{\text{MIP}}$ & CL4SRec & DuoRec & MCLSR & Improv. \\ 
\hline
\multirow{6}{*}{Books}  & Recall@20          & 1.368  & 3.787 & 6.274 & 5.489 & 5.721 & 6.336  & 6.544 & \underline{6.838} & \textbf{7.469$^\dagger$} &  9.2\%  \\
                        & NDCG@20          & 0.597  & 1.923 & 2.825 & 2.262  & 2.706 & 2.964 & 3.161 & \underline{3.257} & \textbf{3.479$^\dagger$} &  6.8\%  \\
                        & Hit@20        & 3.013  & 8.710 & 12.765 & 11.402 & 11.730  & 13.052 & 13.520 & \underline{14.173} & \textbf{15.542$^\dagger$} &  9.6\%  \\
                        & Recall@50          & 2.400  & 6.335 & 9.349 & 8.467  & 8.455 & 9.684 & 10.240 & \underline{10.826} & \textbf{11.583$^\dagger$} &  6.9\%  \\
                        & NDCG@50          & 0.826  & 2.600 & 3.627 & 3.082  & 3.434 & 3.894 & 4.113 & \underline{4.308} & \textbf{4.647$^\dagger$} & 7.9\%   \\
                        & Hit@50        & 5.219  & 13.597 & 18.547 & 17.202  & 16.865 & 19.142 & 20.170 & \underline{21.366} & \textbf{23.042$^\dagger$} & 7.8\%   \\
\hline
\multirow{6}{*}{Clothing} & Recall@20          & 1.200  & 1.623 & 2.646 & 1.678 & 2.242 & 2.704 & 2.863 & \underline{2.940} & \textbf{3.138$^\dagger$} & 6.7\%   \\
                        & NDCG@20          & 0.374  & 0.559 & 0.854 & 0.427 & 0.659 & 0.873 & 0.927 & \underline{1.018} & \textbf{1.081$^\dagger$} & 6.2\%   \\
                        & Hit@20        & 2.139  & 2.777 & 4.188 & 3.467 & 3.684 & 4.343 & 4.467 & \underline{4.829} & \textbf{5.138$^\dagger$} &  6.4\%  \\
                        & Recall@50          & 2.715  & 2.948 & 4.505 & 2.774 & 3.309 & 4.522 & 4.651 & \underline{4.956} & \textbf{5.352$^\dagger$} & 7.9\%    \\
                        & NDCG@50          & 0.640  & 0.778 & 1.151 & 0.723 & 0.829 & 1.116 & 1.199 & \underline{1.356} & \textbf{1.464$^\dagger$} &  8.0\%  \\
                        & Hit@50        & 4.833  & 5.085 & 6.705 & 5.052 & 5.812 & 6.723 & 7.155 & \underline{7.785} & \textbf{8.503$^\dagger$} &  9.2\% \\
\hline
\multirow{6}{*}{Toys}     & Recall@20        & 0.928  & 3.214 & 6.343 & 5.315 & 6.593 & 6.670 & 6.983 & \underline{7.841} & \textbf{8.254$^\dagger$} & 10.3\% \\
                        & NDCG@20          & 0.510 & 1.641 & 2.912 & 2.114 & 2.817 & 3.073 & 3.072 & \underline{3.418} & \textbf{3.726$^\dagger$} & 9.0\%   \\
                        & Hit@20        & 2.496 & 6.926 & 12.838 & 11.075 & 13.153 & 13.474 & 14.079 & \underline{15.331} & \textbf{16.661$^\dagger$} & 8.7\%    \\
                        & Recall@50          & 1.844 & 5.406 & 10.264 & 8.962 & 10.018 & 10.730 & 11.300 & \underline{12.463} & \textbf{13.328$^\dagger$} & 6.9\%   \\
                        & NDCG@50          & 0.774 & 2.216 & 3.899 & 2.952 & 3.690 & 4.072 & 4.095 & \underline{4.612} & \textbf{5.081$^\dagger$} &  10.2\% \\
                        & Hit@50        & 4.760 & 11.554 & 19.837 & 17.282 & 19.400 & 20.363 & 21.330 & \underline{23.389} & \textbf{25.462$^\dagger$} & 8.9 \%  \\
\hline
\multirow{6}{*}{Gowalla} & Recall@20          & 1.206  & 5.642 & 8.581 & 5.559 & 7.869 & 7.823 & 8.804 & \underline{8.973} & \textbf{9.317$^\dagger$} &  3.8\%  \\
                        & NDCG@20          & 1.191  & 5.536 & 7.546 & 3.891 & 6.819 & 7.351 & 7.601 & \underline{7.618} & \textbf{7.759$^\dagger$} & 1.9\% \\
                        & Hit@20        & 5.874  & 22.450 & 28.931 & 19.052  & 26.315 & 27.676 & 29.853 & \underline{30.075} & \textbf{31.832$^\dagger$} &  5.8\% \\
                        & Recall@50          & 2.084  & 9.623 & 13.838 & 9.891  & 12.793 & 12.710 & 14.372 & \underline{15.195} & \textbf{15.972$^\dagger$} & 5.1\%   \\
                        & NDCG@50          & 1.678  & 7.784 & 10.510 & 5.725 & 9.107 & 9.752 & 10.630 & \underline{10.735} & \textbf{11.012$^\dagger$} &  2.6\% \\
                        & Hit@50        & 9.716  & 34.321 & 42.380 & 32.041 & 38.613 & 39.463 & 43.659 & \underline{44.618} & \textbf{46.217$^\dagger$} &  3.5\%   \\
\toprule[0.8pt]
\end{tabular}
\label{tab:results}
\end{table*}
}

\subsection{Performance Comparsion}

The experimental results of our method with state-of-the-art baselines are listed in Table \ref{tab:results}, from where we have the following key findings:
\begin{itemize}[leftmargin=*]
\item The performance of POP is the worst since it directly uses rudimentary statistical methods to recommend the most frequently occurring items in the training data, which fails to learn the preference of users. GRU4Rec outperforms POP on four datasets, demonstrating the effectiveness of neural networks for SR. However, GRU4Rec performs poorly compared to other neural network-based methods. It shows that directly using the representation of the last step of RNN is not enough for SR, which is due to the forgetting problem of RNN. As such, dependencies between items cannot be effectively extracted.
\item We can observe that SASRec and ComiRec-SA surpass GRU4Rec on four datasets, which demonstrates the strength of the multi-head self-attention mechanism for SR. It may benefit from two aspects: First, the attention mechanism can capture the long-term dependencies within the sequence. Second, it will assign more significant weight to more important items, which filters the noise in the sequence.
\item By comparing GCSAN and ComiRec-SA, we can observe the benefits brought by graph neural networks. It is because graph neural networks can capture more complex item transitions by converting transitions within sequences to graphs, resulting in better performance.
\item S$^3$-Rec$_{\text{MIP}}$ and CL4SRec exhibit relatively good performance among baseline methods, indicating the significance of contrastive learning for SR. It is because the contrastive mechanism can learn the extra self-supervised signal for SR, which alleviates the data sparsity problem. However, both two methods perform data-level augmentation on each sequence. Due to the less information diversity between the contrastive pairs, the self-supervised signal will be too weak to learn informative embedding.
\item DuoRec surpasses S$^3$-Rec$_{\text{MIP}}$ and CL4SRec in most cases. The reason is that DuoRec applies model-level augmentation (\ie dropout) to the sequence and performs contrastive learning with a positive sampling strategy, which enhances the diversity of information for contrastive learning. However, the information in each sequence is limited, which means it can obtain limited self-supervised signals without exploring the rich collaborative information and co-action information.
\item MCLSR significantly outperforms all baselines overall four datasets consistently. Specifically, the average improvement of MCLSR over the best baseline is $7.0\%$ on four datasets, demonstrating its effectiveness for SR. Different from previous contrastive learning-based methods, the proposed MCLSR leverages a multi-level contrastive mechanism and extracts the complex collaborative information and co-action information for SR, which learns discriminative user and item embeddings.
\end{itemize}

\begin{figure*}[t]
  \centering
  \includegraphics[width=\linewidth, angle=0]{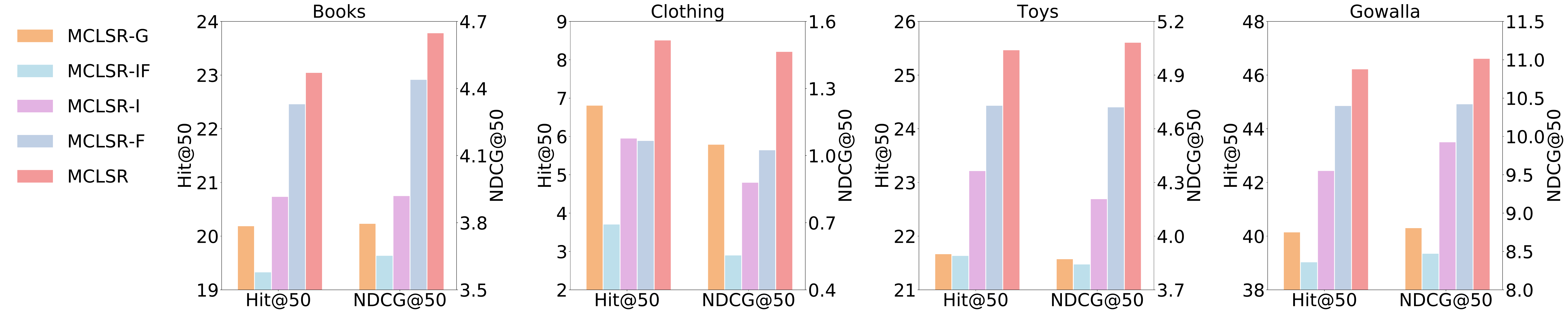}
  \caption{Ablation study on four datasets.}
  \label{fig:ablationStudy}
\end{figure*}


\begin{figure}[t]
    \centering
    \subfloat[Sensitive of parameter $\alpha$]{
    \centering
    \includegraphics[width=\linewidth]{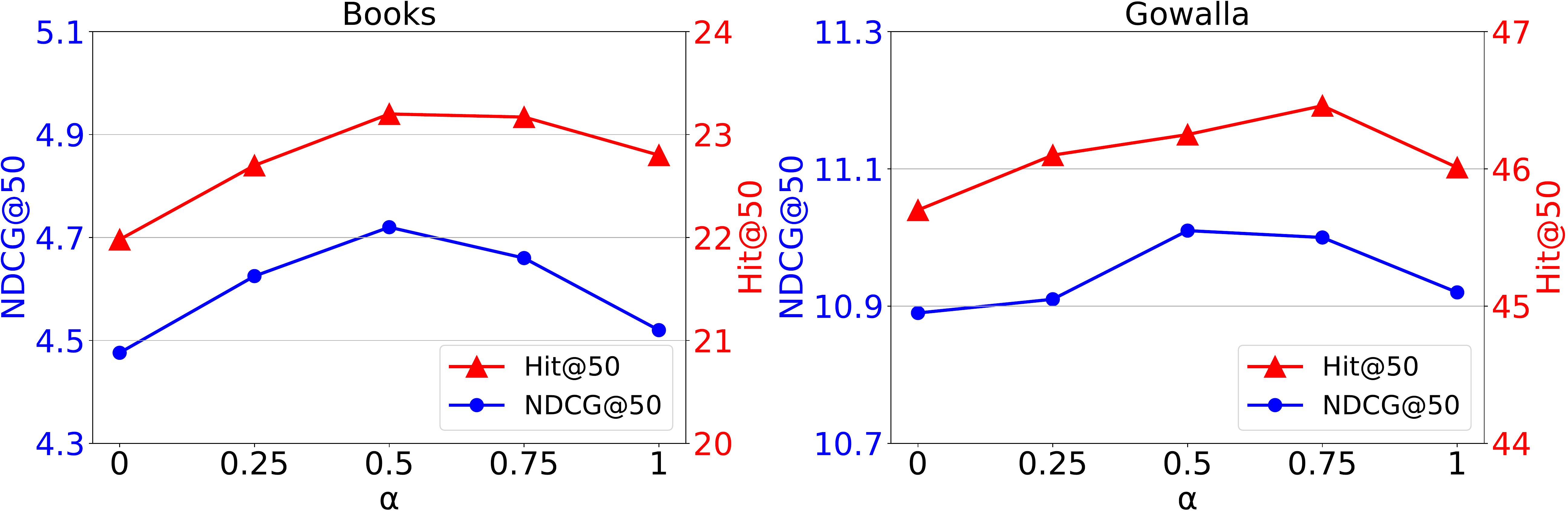}
    \label{fig:impactAlpha}
    }
    
    \subfloat[Sensitive of parameter $\beta$]{
    \centering
    \includegraphics[width=\linewidth]{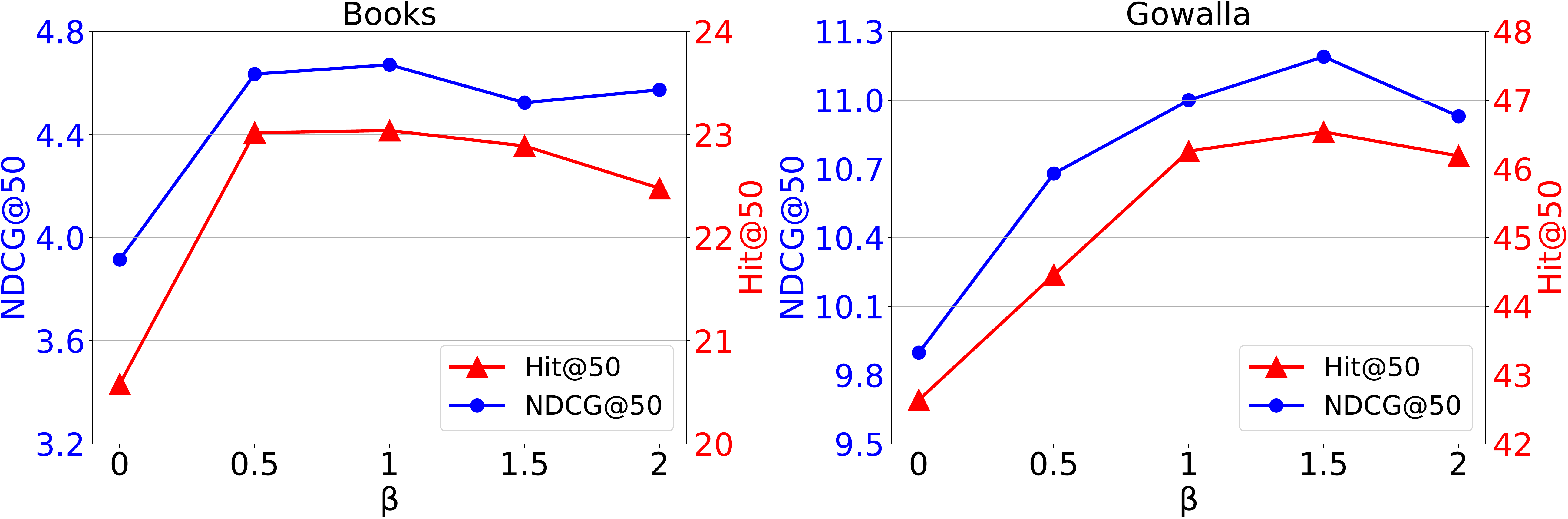}
    \label{fig:impactBeta}
    }
    
    \subfloat[Sensitive of parameter $\gamma$]{
    \centering
    \includegraphics[width=\linewidth]{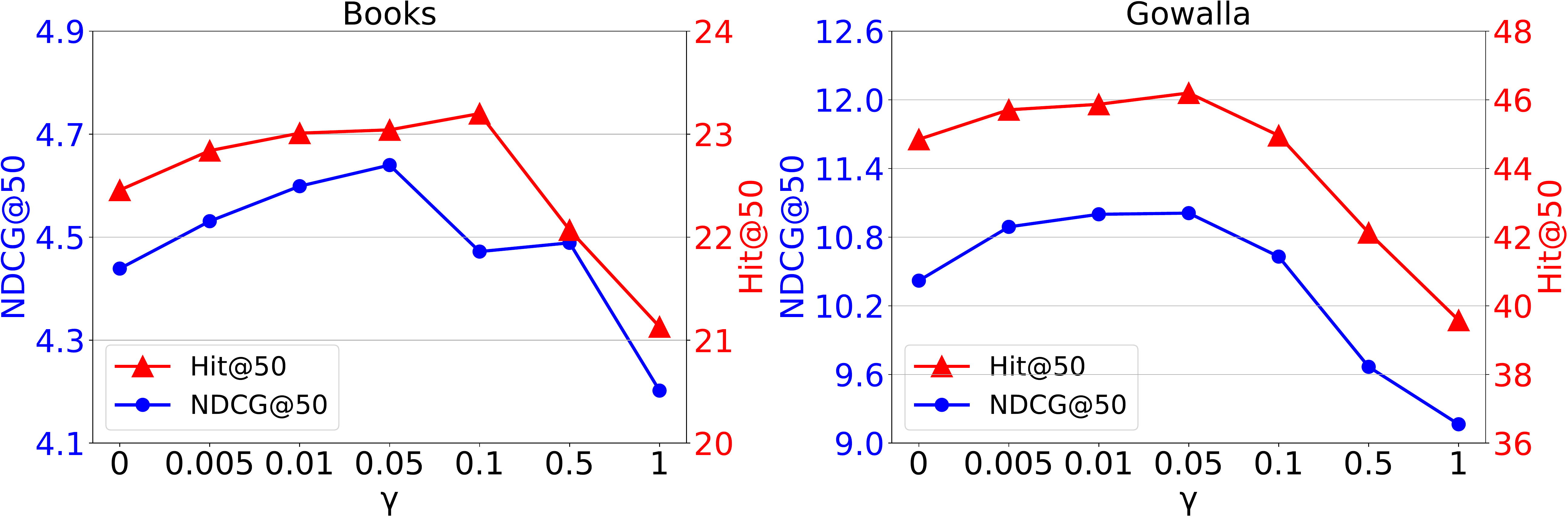}
    \label{fig:impactGamma}
    }
    
    \caption{Impact of trade-off parameter.}
    \label{fig:impactParameter}
\end{figure}


{
\renewcommand{\arraystretch}{1}
\begin{table*}[!t]
    \setlength{\tabcolsep}{2.5pt}
    \small
    \centering
    \caption{The performance of MCLSR with varied depth of GNN layers in terms of Metrics@50.}
    \begin{tabular}{clccclccclccclccc}
    \toprule[0.8pt]
    \multirow{2}{*}{Depth} & & \multicolumn{3}{c}{Books} & & \multicolumn{3}{c}{Clothing} & & \multicolumn{3}{c}{Toys} & & \multicolumn{3}{c}{Gowalla}
    \\ 
    \cline{3-5} \cline{7-9} \cline{11-13} \cline{15-17}
    & & Recall  & NDCG & Hit Rate & & Recall & NDCG & Hit Rate & & Recall & NDCG & Hit Rate & & Recall & NDCG & Hit Rate     \\
    \hline
    $l$ = 0   & & 9.881 & 3.799 & 20.032 & & 13.025 & 8.805 & 40.141 & & 11.052 & 3.847 & 21.673 & & 13.265  & 8.728 & 40.273 \\
    $l$ = 1   & & 10.853 & 4.166 & 21.782 & & 15.897 & 10.889 & 44.543 & & 13.165 & 4.744 & 25.069 & & 15.769 & 10.743 & 45.620 \\
    $l$ = 2   & & \textbf{11.583} & \textbf{4.647} & \textbf{23.042} & & \textbf{15.972} & \textbf{11.012} & \textbf{46.217} & & 13.328 & 5.081 & 25.462 & & \textbf{15.972} & \textbf{11.012} & \textbf{46.217}    \\
    $l$ = 3   & & 10.085 & 3.936 & 20.547 & & 15.021 & 10.042 & 43.775 & & \textbf{13.720} & \textbf{5.187} & \textbf{26.417} & & 14.948 & 10.094 & 43.665     \\
    \toprule[0.8pt]
    \end{tabular}
    \label{tab:depth}
\end{table*}
}

\subsection{Ablation Study}
As the proposed MCLSR outperforms all kinds of baselines consistently, we investigate the effectiveness of critical components to analyze the proposed method deeply and comprehensively. Specifically, we conduct ablation studies by
comparing four variants with the complete model on four datasets. (1)``MCLSR-G'' represents removing the graph encoder layer and directly using the initial embedding of user and item. (2) ``MCLSR-IF'' means removing two levels of contrastive mechanism and only optimizing the loss function for target prediction. (3) ``MCLSR-F'' directly removes the feature-level contrastive mechanism. (4) ``MCLSR-I'' represents removing the interest-level contrastive mechanism. From the results in Figure \ref{fig:ablationStudy}, we can make the following observations:
\begin{itemize}[leftmargin=*]
\item There is a significant drop when removing the graph encoder layer~(\ie MCLSR-G). It verifies the significance of the collaborative information hidden in the user-item graph and the co-action information hidden in the user-user/item-item graph. Besides, it also demonstrates the effectiveness of the graph encoder layer.
\item The performance of MCLSR would be remarkably decreased when removing two levels of contrastive mechanism~(\ie MCLSR-IF), indicating the crucial role of two levels of contrastive mechanism in learning the extra self-supervised signals and alleviating the data sparsity problem.
\item Compared with MCLSR, MCLSR-I obtains dramatically worse results on four datasets. It is because MCLSR-I losses the self-supervised signal obtained by contrasting the user-item view and sequential view. The results demonstrate the importance of interest-level contrastive mechanisms to learn significant self-supervised signals.
\item MCLSR-F is inferior to the complete model MCLSR, especially on Clothing datasets, indicating the importance of co-action information in learning the informative embeddings of users and items from user-user and item-item graphs.
\end{itemize}


\begin{figure*}[tb]
    \centering
    \subfloat[SASRec.]{
    \centering
    \includegraphics[height=0.18\linewidth]{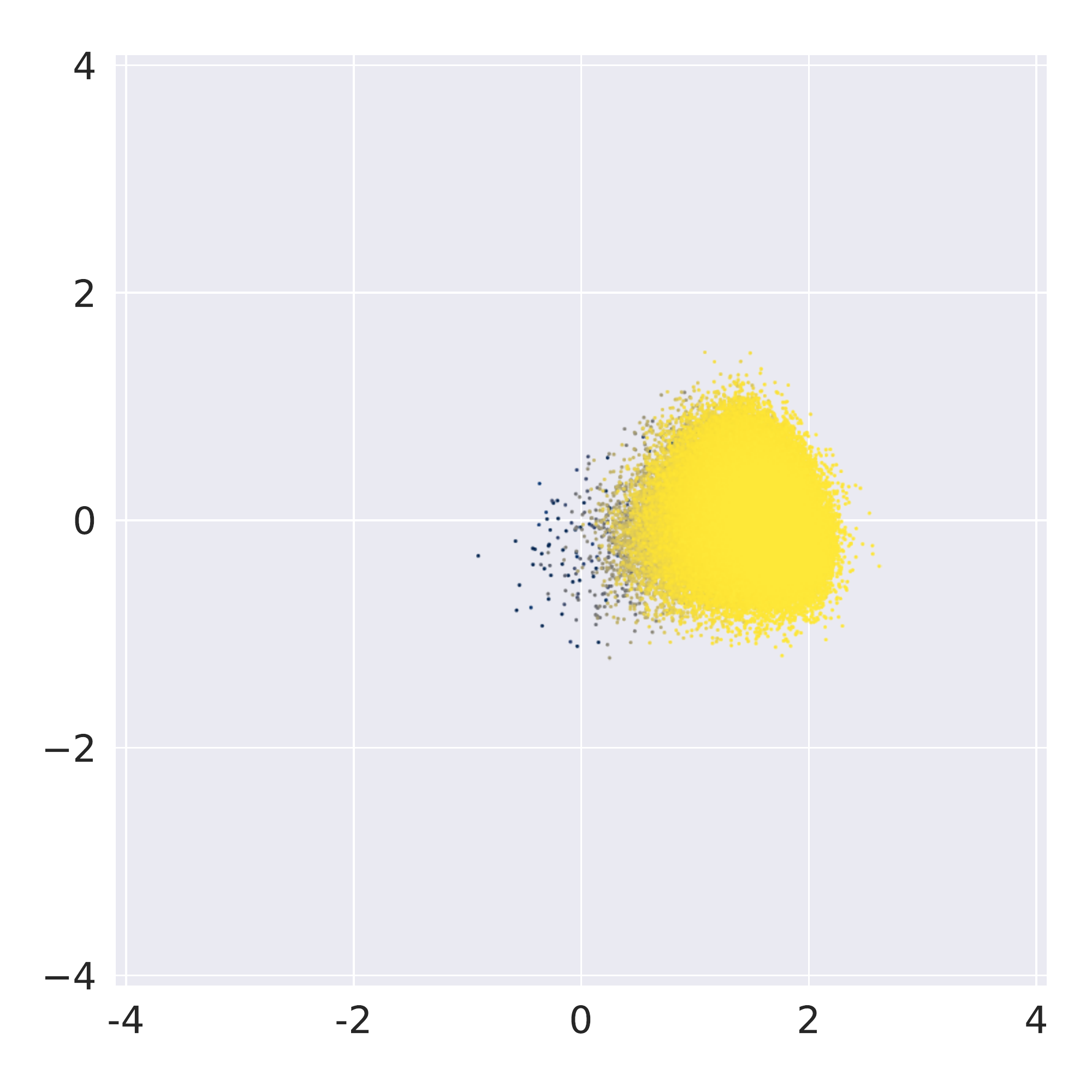}
    \label{fig:1}
    }
    \subfloat[CL4SRec.]{
    \centering
    \includegraphics[height=0.18\linewidth]{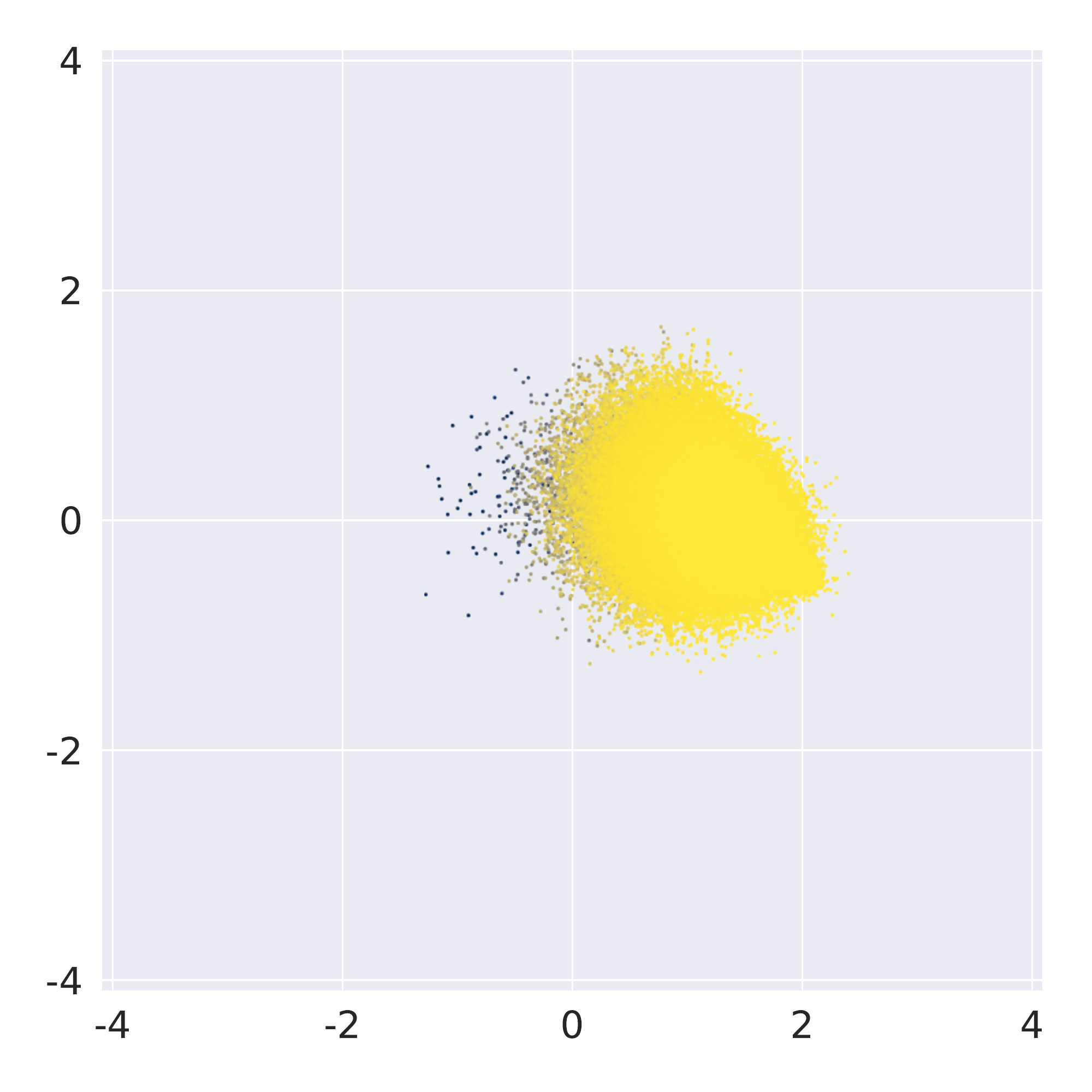}
    \label{fig:2}
    }
    \subfloat[DuoRec.]{
    \centering
    \includegraphics[height=0.18\linewidth]{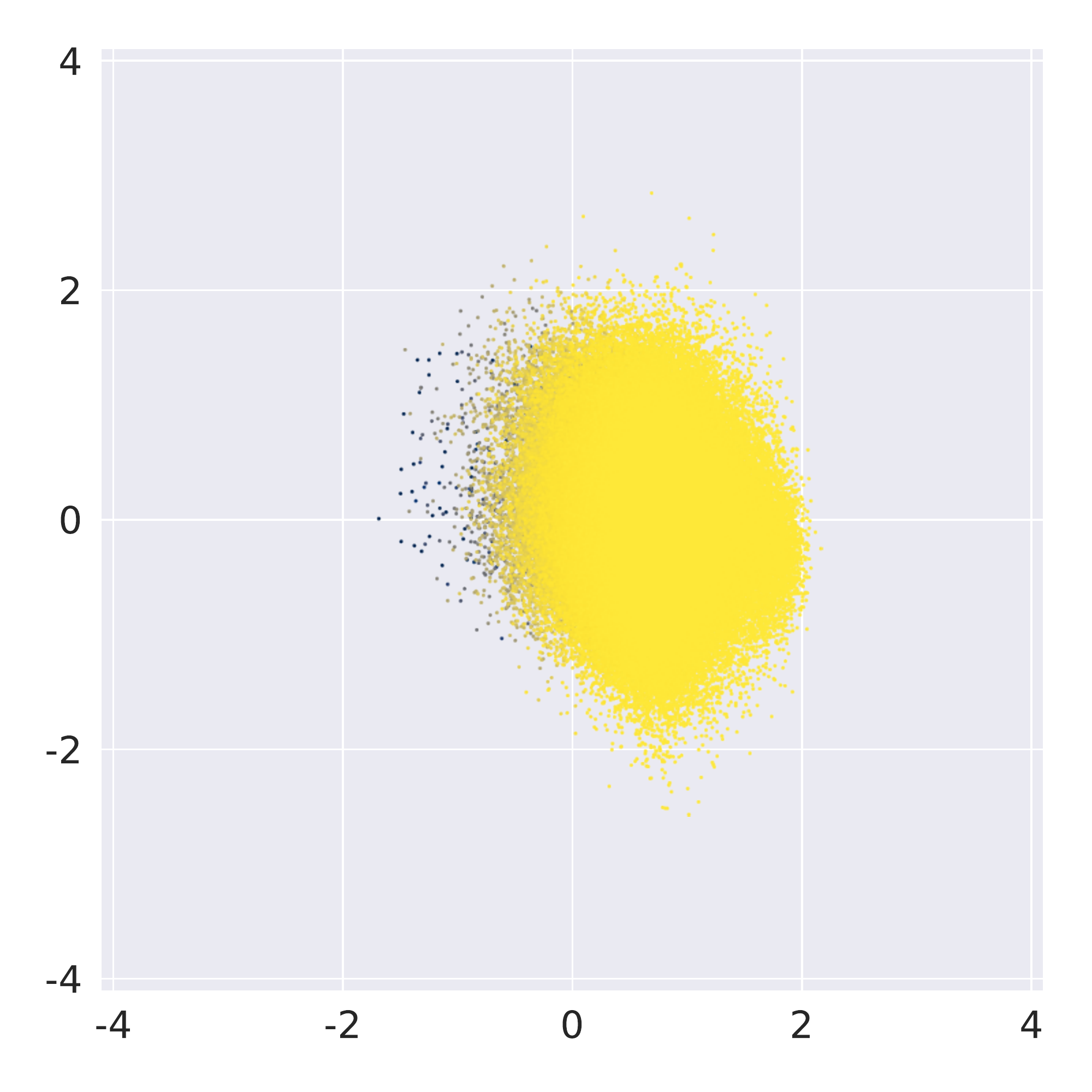}
    \label{fig:3}
    }
    \subfloat[MCLSR w/o CL.]{
    \centering
    \includegraphics[height=0.18\linewidth]{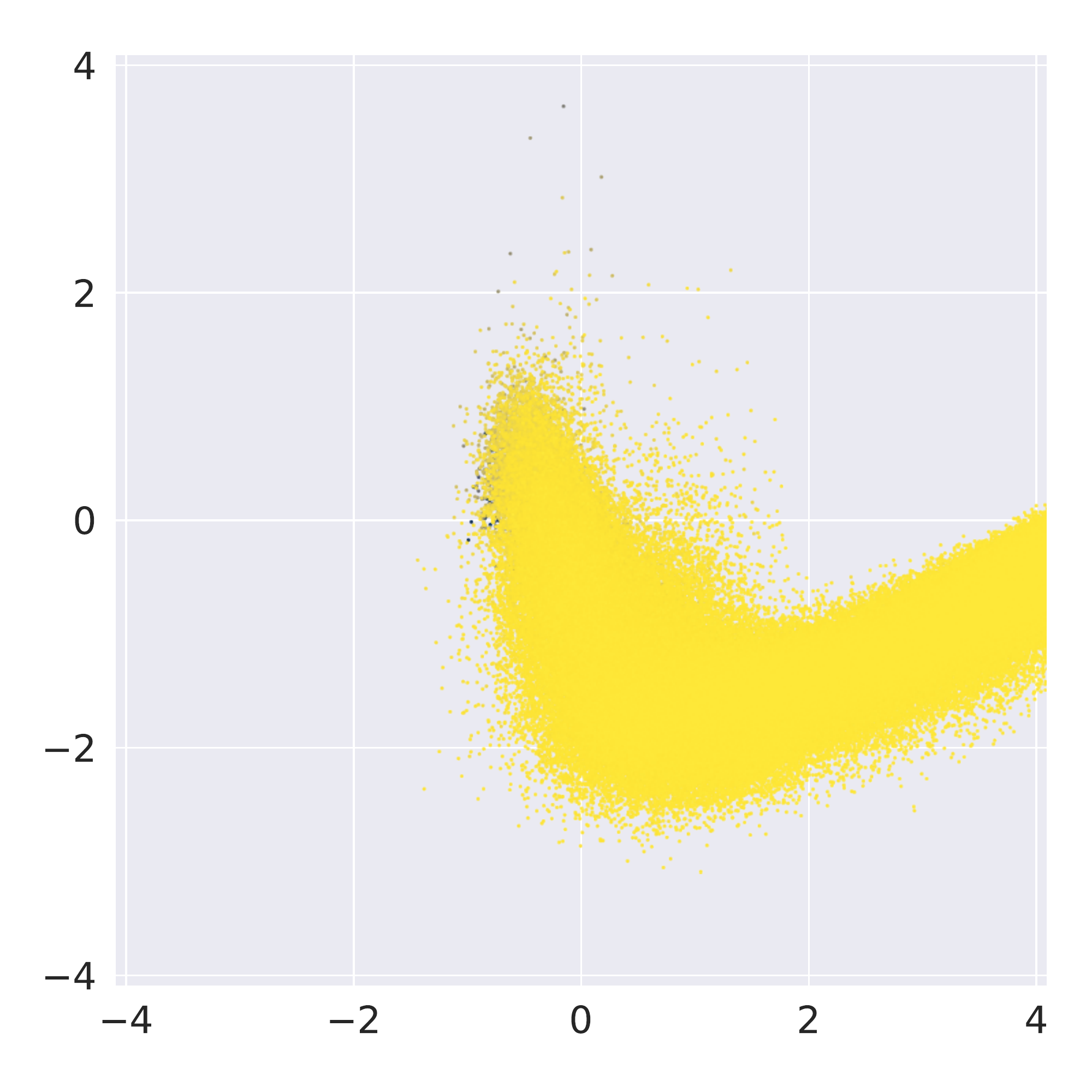}
    \label{fig:4}
    }
    \subfloat[MCLSR.]{
    \centering
    \includegraphics[height=0.18\linewidth]{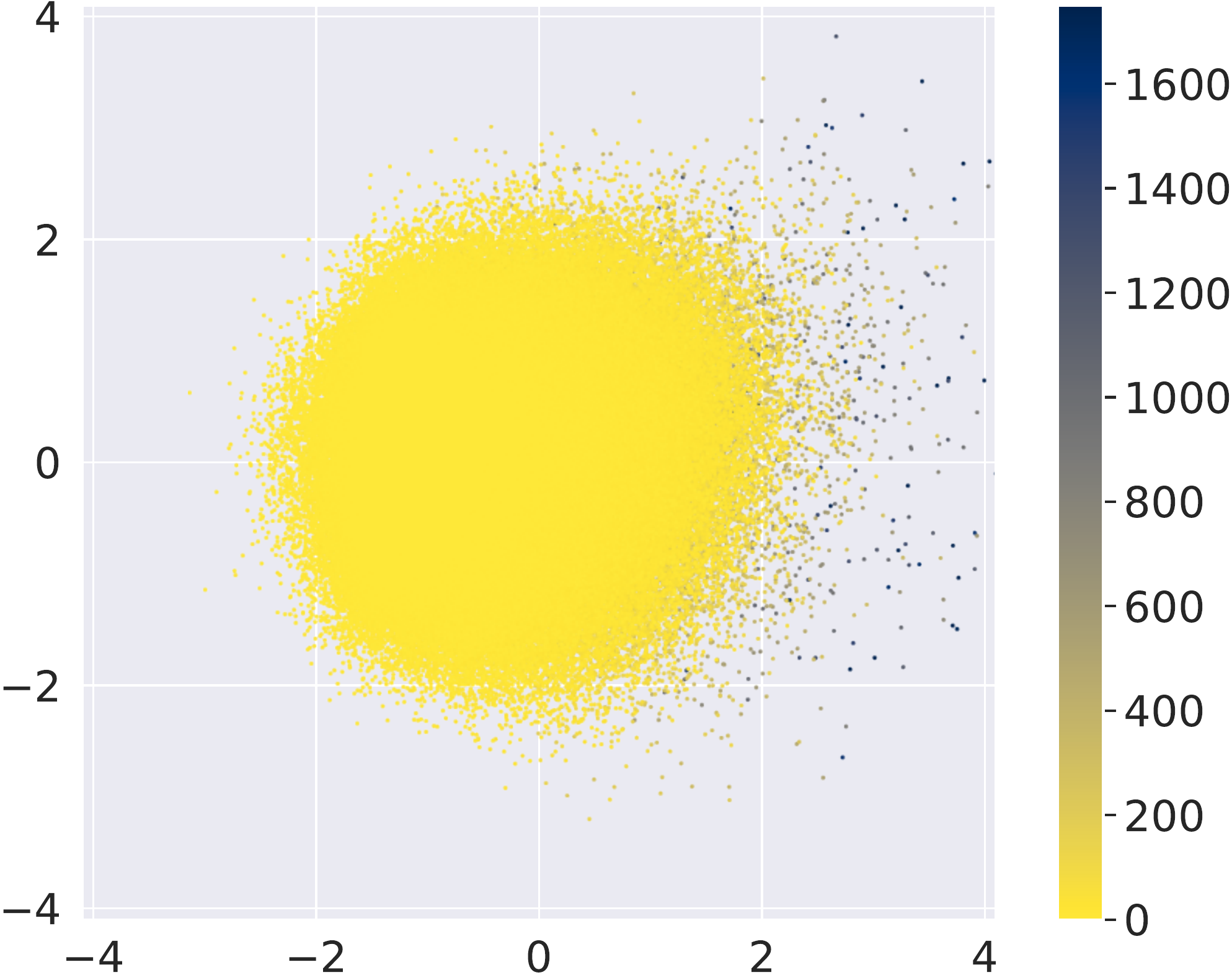}
    \label{fig:5}
    }
    \caption{Item embeddings of selected methods on Book dataset.}
    \label{fig:embedding}
\end{figure*}

\subsection{Parameter Sensitive}
Here, we investigate the impact of important hyper-parameter settings on the performance of MCLSR, including trade-off parameters $\alpha$, $\beta$, $\gamma$, and the number of propagation layers in GNN. The results are shown in Figure \ref{fig:impactParameter} and Table \ref{tab:depth}

\paratitle{Impact of the parameter $\alpha$.}
The trade-off parameter $\alpha$ in Equation \ref{eq:finalInterest} controls the proportion of general interest and current interest during train process. It can be seen from Figure \ref{fig:impactAlpha} that MCLSR performs worst when $\alpha$ is set to $0$, which is due to the inconsistency between the training process and the inference process. Besides, the performance dramatically degrades when $\alpha$ is set to $1$, indicating the importance of general interest. From the results, MCLSR obtains the best performance when $\alpha = 0.25$ on Books and $\alpha=0.75$ on Gowalla. 

\paratitle{Impact of the parameter $\beta$.}
The trade-off parameter $\beta$ determines the influence of the interest-level contrastive mechanism during the training process.
From Figure \ref{fig:impactBeta}, it can be observed that the performance of MCLSR shows a significant rise when $\beta$ increases from $0$ to $0.5$, which demonstrates the crucial role of the interest-level contrastive mechanism.
Besides, the results of MCLSR become slightly worse when $\beta > 1.5$ on Gowalla, which means excessive attention to collaborative information may hurt performance. 

\paratitle{Impact of the parameter $\gamma$.}
The trade-off parameter $\gamma$ controls the influence of the feature-level contrastive mechanism.
From the results in Figure \ref{fig:impactGamma}, MCLSR obtains better performance when $\gamma$ increases from $0$ to $0.05$ on two datasets, which demonstrates the effectiveness of feature-level contrastive learning.
Besides, the performance of the model drops significantly when $\gamma$ is set to 1, which shows that focusing too much on the co-action signals of users and items will deteriorate the performance of the model. 

\paratitle{Impact of the depth of GNN layers.} 
To deeply investigate whether MCLSR benefits from the graph information, we search the number of graph encoder layers $l$ in the range of $\{0, 1, 2, 3\}$ and summarize the results in Table \ref{tab:depth}. It can be observed that:
\begin{itemize}[leftmargin=*]
\item The information on the user-item graph, user-user graph, and item-item graph is significant for SR. Specifically, MCLSR with $l=0$ obtains dramatically worse results, and there is a significant improvement when setting $l=1$ for MCLSR.
\item Increasing the depth of GNN is able to enhance the predictive results. More specifically, MCLSR with $l=2$ performs better than MCLSR with $l=1$ on four datasets and MCLSR achieves the best performance on Toys when $l=3$, which indicates that higher order of propagation obtains more effective collaborative information from three graph views.
\item Higher layer of GNN may deteriorate the performance of MCLSR. Specifically, MCLSR with $l=3$ performs worse than MCLSR with $l=2$ in most cases, which may be due to the overfitting problem of GNNs~\cite{wang2019neural}. 
\end{itemize}

    
    

    

\subsection{Qualitative Analysis}
Except for performance scores, we also provide qualitative results to demonstrate the superiority of MCLSR further. Specifically, we project the learned item embedding into two-dimensional space by SVD~\cite{qiu2022contrastive} and show the learned space of selected methods on the Book dataset in Figure \ref{fig:embedding}.

Form Figure \ref{fig:1} we can observe that the item embedding learned by SASRec degenerated into a narrow cone. According to~\cite{gao2018representation,wang2019improving} such phenomena deteriorate the model’s capacity as the learned embedding does not have enough capacity to model the diverse features. 
Comparing Figure \ref{fig:2}, \ref{fig:3} and \ref{fig:1}, the learned embedding spaces of CL4SRec and DuoRec are better than SASRec. It is because CL4SRec and DuoRec devise auxiliary self-supervised objectives for data representation learning based on data-level augmentation and model-level augmentation, respectively. However, directly exploiting the self-supervised signals from the sequence is insufficient for SR.
In contrast, 
Figure \ref{fig:4} and \ref{fig:5} show that the learned embeddings of MCLSR without and with CL. It can be observed that the learned embeddings of MCLSR~(Figure \ref{fig:5}) are somewhat uniformly distributed around the origin and not strictly in a narrow cone, which effectively expands the embedding space and has more capacity to model the diverse features of items. We argue that this is because MCLSR learns
the representations of users and items through a cross-view contrastive learning paradigm on two levels. Specifically, the interest-level contrastive mechanism jointly learns the collaborative information and the sequential transition patterns, and the feature-level contrastive mechanism captures the co-action signals when learning the user and item features. In this way, MCLSR obtains discriminative item and user representations without extra labels.

\section{Conclusion}

This study presents a multi-level contrastive learning framework for sequential recommendation.
Different from previous methods, we design four informative views
~(\ie sequential view, user-item view, user-user view, and item-item view) 
and learn the self-supervised signals via cross-view contrastive learning at two different levels.
The interest-level contrastive mechanism learns the complementary information from collaborative information and sequential transition patterns and the feature-level contrastive mechanism mines the co-action information between users and items.
Comprehensive experiments demonstrate that the proposed method significantly outperforms baselines over four datasets, 
indicating it has excellent potential to solve real-world recommendation problems.

\section*{Acknowledgments}
This work was supported in part by the National Natural Science Foundation of China under Grant No.61602197, Grant No.L1924068, Grant No.61772076, in part by CCF-AFSG Research Fund under Grant No.RF20210005, and in part by the fund of Joint Laboratory of HUST and Pingan Property \& Casualty Research (HPL). 

\bibliographystyle{ACM-Reference-Format}
\bibliography{acmart}

\end{document}